\documentclass[preprint,aps,showpacs,nofootinbib,preprintnumbers,amsmath,amssymb]{revtex4-1}
\usepackage{amsfonts}
\usepackage{amssymb}
\usepackage{subfigure,amsfonts, latexsym}
\usepackage[notcite,color,final]{showkeys}
\usepackage{epsfig}
\usepackage{graphicx}% Include figure files
\usepackage{subfigure}
\usepackage{dcolumn}% Align table columns on decimal point
\usepackage{bm}% bold math
\usepackage[usenames ,dvipsnames]{xcolor}
\usepackage{slashed}
\begin{document}
\title{Correlation Between Muon $g-2$ and $\mu\rightarrow{e}{\gamma}$}

\author{Wei-Chi~Chiu$^{1}$\footnote{ericgreat@gmail.com}, Chao-Qiang Geng$^{1,2,3}$\footnote{geng@phys.nthu.edu.tw},
Da~Huang$^{1}$\footnote{dahuang@phys.nthu.edu.tw}}
\affiliation{$^{1}$Department of Physics, National Tsing Hua University, Hsinchu, Taiwan\\
  $^{2}$Physics Division, National Center for Theoretical Sciences, Hsinchu, Taiwan\\
$^{3}$Chongqing University of Posts \& Telecommunications, Chongqing, 400065, China
}

\date{\today}

\begin{abstract}
While the muon $g-2$ anomaly can be successfully explained by some new physics models, most of them are severely constrained
by the $\mu \to e \gamma$ bound. This tension is more transparent from the effective field theory perspective, in which the two phenomena
are encoded in two very similar operators. However, with the ${\cal O}(1)$ Wilson coefficients, the current upper bound on $\mu \to e \gamma$
indicates a new-physics cutoff scale being five orders smaller than that needed to eliminate the $(g-2)_\mu$ anomaly. By summarizing all the formulae
from the one-loop contributions to the muon $g-2$ with the internal-particle spin not larger than 1, we point out two general methods to reconcile
the conflict between the muon $g-2$ and $\mu \to e \gamma$: the GIM mechanism and the non-universal couplings. For the latter method,
we use a simple scalar leptoquark model as an illustration.
\end{abstract}

\maketitle

%\pacs{
%\newpage

\section{Introduction}\label{intro}
The Standard Model (SM) provides an excellent description of elementary particles and interactions,
which has further been confirmed by the discovery of the SM-like Higgs particle at the Large Hadron Collider
(LHC)~\cite{Beringer:1900zz, Aad:2012tfa, Chatrchyan:2012ufa}. Nevertheless, there are still some experimental
results showing the tantalizing hints to new physics beyond the SM. Especially, one of the biggest discrepancies between
the experimental values and the SM predictions comes from the muon anomalous magnetic moment (AMM) or $(g-2)_\mu$ for short,
known as the  $(g-2)_\mu$ anomaly.

In 2001, the E821 experiment at Brookhaven National Lab (BNL)~\cite{Bennett:2004pv, Bennett:2006fi} showed that
the measured value of $a_\mu\equiv (g-2)_\mu/2$ exceeds the SM prediction by about $2.4\sim3.6\sigma$, where the different standard deviations
are originated from the various theoretical methods for estimating the hadronic contributions, by applying either
the $e^{+}e^{-}\rightarrow hadrons$~\cite{Davier:2010nc} or the $\tau$-based~\cite{Alemany:1997tn} data.
% as well as $e^{+}e^{-}\rightarrow hadrons$ \cite{Davier:2010nc} or $\tau$-based data \cite{Alemany:1997tn}.
The value of the discrepancy between the measurement and SM prediction is given by~\cite{Beringer:1900zz, Bennett:2006fi, Kinoshita:2005sm, Czarnecki:2002nt, Hagiwara:2011af, Davier:2004gb, Passera:2004bj}
\begin{equation}
\label{eq:muon g-2}
\Delta { a }_{ \mu  }={ a }_{ \mu  }^{\rm exp }-{ a }_{ \mu  }^{\rm SM }
 = { { 287(63)(49)\times 10 }^{ -11 } }\, .
\end{equation}
To explain this anomaly, many models have been proposed (see~\cite{Jegerlehner:2009ry} and references therein). 
For example, by including the mixings of the muon
with some new TeV-scale heavy leptons~\cite{Czarnecki:2001pv, Kannike:2011ng}, the new contribution to
$(g-2)_\mu$ is roughly the same order as the $W$- and $Z$-boson ones due to the enhancement from the heavy lepton masses.

However, most of these models predict visible lepton flavor violating (LFV) processes. In particular, the decay of
 $\mu\to e\gamma$ usually gives one of the most stringent constraints, due to the accuracy of its experimental searches.
 For example, the MEG collaboration has recently renewed the upper bound on the branching ratio of this process to
 ${\cal B}(\mu \to e\gamma) \leq  5.7\times10^{-13}$ ($90\% $ C.L.)~\cite{Adam:2013mnn}.
 As a result, it is natural and necessary to use this updated limit to constrain the structure of new physics behind the $(g-2)_\mu$ anomaly.

The main purpose of the present paper is to propose some general methods of building models to not only explain the $(g-2)_\mu$ anomaly
 but also satisfy the stringent $\mu \to e \gamma$ constraint. We first explore this problem from the effective field theory (EFT) perspective,
 finding that it is very challenging to accomplish both processes simultaneously  with the same UV cutoff and the similar order of
 the Wilson coefficients. We then focus on some simplified models with new physics perturbatively coupled to the SM fields
 so that the leading order contributions to $(g-2)_\mu$ and $\mu \to e\gamma$ appear at the one-loop order.
 By collecting all one-loop contributions with different spins ($\leqslant 1$), CP properties and electric charges of new particles,
 we find two general methods to reconcile the tension between the $(g-2)_\mu$ anomaly and the $\mu \to e\gamma$ constraint: 
 the GIM mechanism and the non-universal couplings. For the latter, we present a simple UV-complete leptoquark model as an illustration.
 In our study, we assume that all of the SM contributions to the muon $g-2$ are already well understood and appropriately calculated.
 The discrepancy mainly arises from  new physics. For clarity, we will use $\delta a_{\mu}$ to denote the new physics correction to 
 $(g-2)_\mu$.

This paper is organized as follows. We analyze the correlation between the muon $g-2$ anomaly and $\mu \to e \gamma$
from the EFT perspective in Sec. II. In Sec.~III, we classify the general new physics models by the leading-order one-loop diagrams
for the muon $g-2$ in terms of their different structures and internal particle contents. In Sec.~IV, we proposes two general methods:
the GIM mechanism and the non-universal couplings, which can resolve the tension between these two processes.
%For the latter, a simple leptoquark model is discussed in detail as an example.
Finally, a short summary is given in Sec.~V.

\section{General analysis from effective operators}\label{SecEFT}
It is quite natural to consider the constraint on new physics from the LFV mode $\mu\to e\gamma$
 to explain the $(g-2)_\mu$ anomaly from the EFT perspective since the structure of the leading effective operators for both processes is
 essentially the same except for that the outgoing muon is replaced by an electron, as illustrated in Fig.~\ref{fig:fig1}.
\begin{figure}[t]
\centering
 \subfigure[short for lof][Muon $g-2$]{   \includegraphics[width=0.3\linewidth]{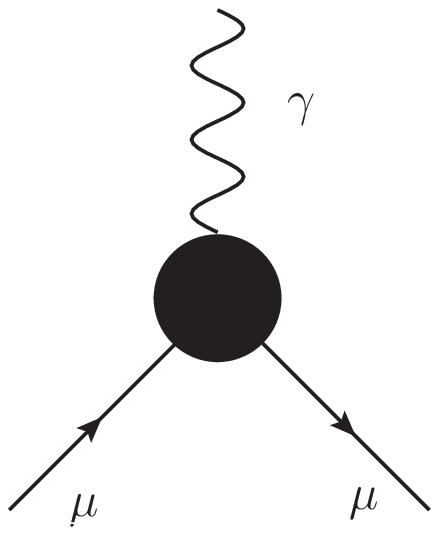}}
   \label{subfig:fig1}
 \hspace{12mm}
 \subfigure[short for lof][ $\mu\rightarrow{e} {\gamma}$]{
   \includegraphics[width=0.3\linewidth]{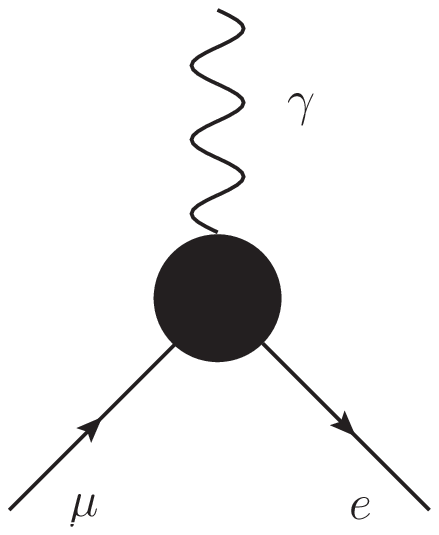}
   \label{subfig:fig2}
}
\caption[short for lof]{Effective Operators for (a) $(g-2)_\mu$ and (b)$\mu\rightarrow{e} {\gamma}$. }
\label{fig:fig1}
\end{figure}
The effective operators for $(g-2)_\mu$ and $\mu \to e\gamma$ are given by:
\begin{eqnarray}
   \delta\mathcal{L}^{a_{\mu}}_{eff} &=& \frac { e }{ \Lambda  } \bar { \mu  } { \sigma  }_{ \mu \nu  }({ C}_{ L } +{ C }_{ R })\mu { F }^{ \mu \nu  },\label{eq:g-2EL}\\
    \delta\mathcal{L}^{\mu\rightarrow{e} {\gamma}}_{eff} &=&  \frac { e }{ \Lambda^\prime  } \bar { e } { \sigma  }_{ \mu \nu  }({ C }^\prime_{ L } { P_{L} } +{ C }^\prime_{ R }{ { P_{R} }  })\mu { F }^{ \mu \nu  },\label{eq:muegammaL}
\end{eqnarray}
respectively, where $ { P_{L} }=(1-{\gamma_{5}})/2$ and ${ P_{R} }=(1+{\gamma_{5}})/2$. At this stage, we differentiate the cutoff scales as $\Lambda$ and $\Lambda^\prime$, and the Wilson coefficients of the left and right-handed couplings as $C_{L,R}$ and $C^{\prime}_{L,R}$  for 
$(g-2)_\mu$ and $\mu \to e\gamma$, respectively. Note that we have extracted the electromagnetic coupling constant $e$ from the Wilson 
coefficients for the normalization.

We can easily derive the contribution to $\delta a_{\mu}$ from Eq.~\eqref{eq:g-2EL}, given by
\begin{align}
\label{}
    \delta { a }_{ \mu  }= { \frac { e({ C }_{ L }+{ C }_{ R } )}{ \Lambda }}/{ \frac { e }{ 2{ m }_{ \mu  } }  }.
\end{align}
With this formula, we find that the desired value of $\delta a_{\mu}\sim 287\times 10^{-11}$ can be obtained by taking $\Lambda \sim { 10 }^{ 8 }$~GeV with the natural value of
$ ({ C }_{ L }+{ C }_{ R } )\sim 1$.

It is also straightforward to evaluate the branching ratio for the decay process $\mu\rightarrow{e} {\gamma}$, given by
\begin{align}
     {\cal B}\left( \mu \rightarrow{e} \gamma  \right) &= \frac{\Gamma \left( \mu\rightarrow{e} {\gamma}  \right) } {\Gamma \left( \mu \rightarrow{e}\nu_\mu \bar { \nu  }_e  \right)} \nonumber\\
    &=\frac { 24{ \pi  }^{ 2 } }{ { { G }_{ F }^{ 2 }m }_{ \mu  }^{ 2 } } \frac { 1 }{ \Lambda^{\prime 2}  } \left( { \left| { C }^\prime_{ L } \right|  }^{ 2 }{ +\left| { C }^\prime_{ R } \right|  }^{ 2 } \right) ,
    \label{eq:Branchingratio}
\end{align}
where $G_{F}$ denotes the Fermi constant and $\Gamma \left( \mu \rightarrow{e}\nu_\mu \bar { \nu  }_e  \right) =   { { G }_{ F }^{ 2 }m }_{ \mu  }^{ 5 } /{ 192{ \pi  }^{ 3 } } $ is used for the normalization. From Eq. \eqref{eq:Branchingratio} and ${\cal B}\left( \mu \rightarrow{e} \gamma  \right)<5.7\times { 10 }^{ -13 }$, one obtains
\begin{align}
\label{}
    \frac { 1 }{ { \Lambda  }^{\prime 2 } } \left( { \left| { C }^\prime_{ L } \right|  }^{ 2 }{ +\left| { C }^\prime_{ R } \right|  }^{ 2 } \right) &\lesssim 3.31\times { 10 }^{ -27 }~{\rm GeV }^{ -2 }.
\end{align}
If we take the Wilson coefficients $C^\prime_{L,R}$ to be $\mathcal{O}(1)$, the cutoff scale $\Lambda^\prime$ should be at least of ${\cal O}(10^{13})$~GeV.

From the above discussion, we see that there is at least a five-order scale gap between the cutoff needed to solve the $(g-2)_\mu$ anomaly
and that required by the $\mu \to e\gamma$ constraint. In other words, if we assume that the same new physics contributes to
both processes, {\it i.e.}, $\Lambda^{'}\sim\Lambda\sim 10^8$ as indicated by the $(g-2)_\mu$ anomaly, the predicted branching ratio
for the $\mu \to e \gamma$ decay is naturally 5 orders larger than the current experimental bound, unless we choose
the Wilson coefficients $C^\prime_{L,R}$ of ${\cal O}(10^{-5})$ or even smaller, which are obviously quite unnatural
from the general EFT philosophy. Clearly, we come to the conclusion that it is challenging for a natural model
to obtain the required contribution to $a_\mu$ while satisfying the current $\mu \to e\gamma$ constraint from the EFT aspect.

%%%%%%%%%%%%%%%%%%%%%%%%%%%%%%%%%%%%%%%%%%%%%%%%%%%%%%%%%
\section{General Results of $(g-2)_\mu$ from One-Loop Diagrams}
\begin{figure}[t]
\begin{center}
\subfigure[]{
\includegraphics[width=0.4\textwidth]{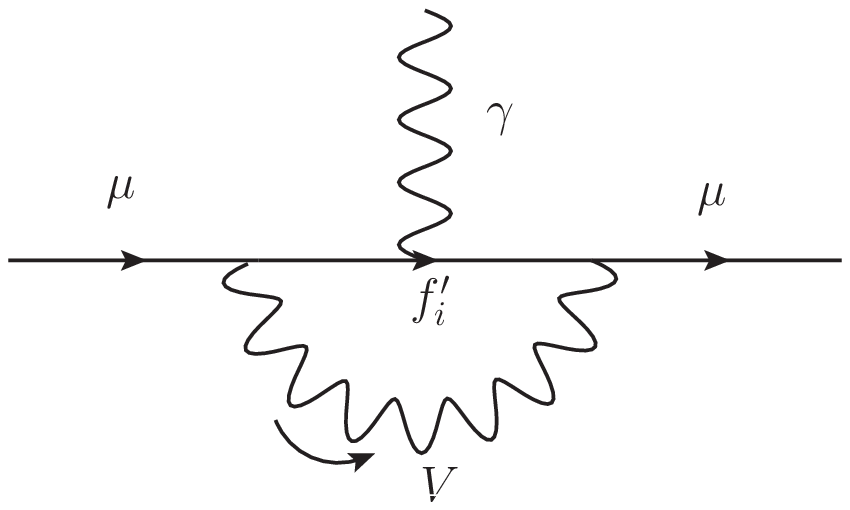}
\label{subfig:NVff}}
\subfigure[]{
\includegraphics[width=0.4\textwidth]{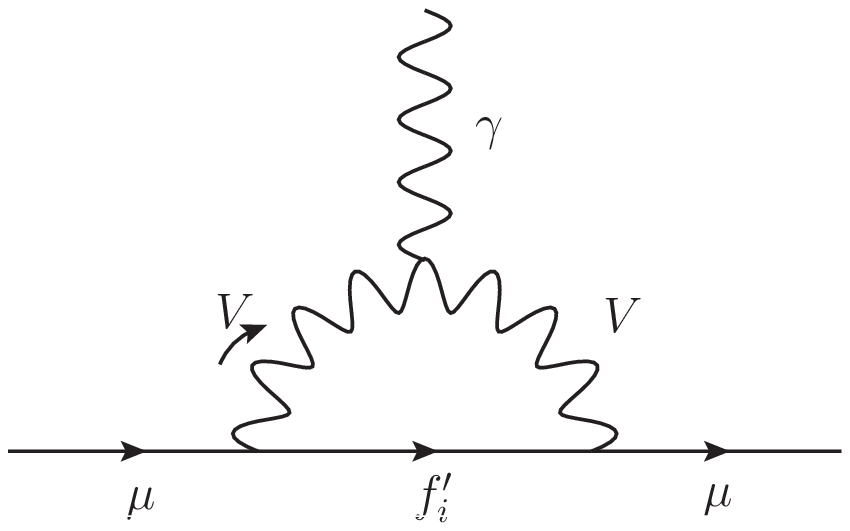}
\label{subfig:CVff}}
\subfigure[]{
\includegraphics[width=0.4\textwidth]{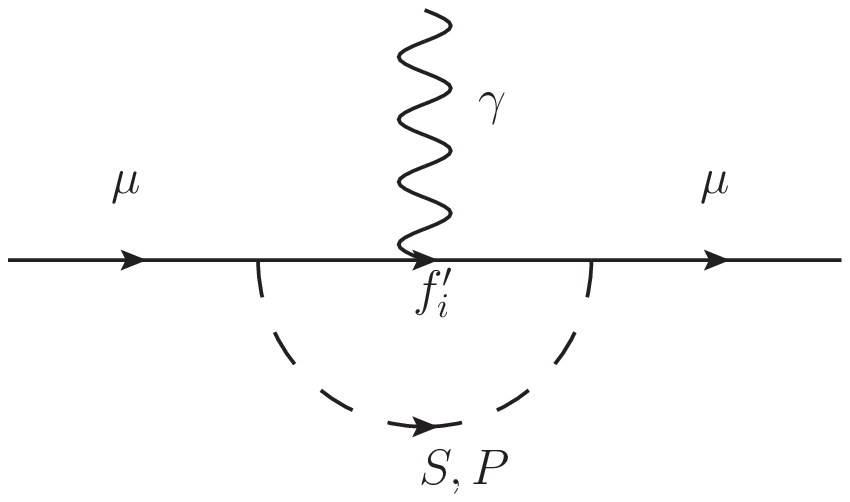}
\label{subfig:NSff}}
\subfigure[]{
\includegraphics[width=0.4\textwidth]{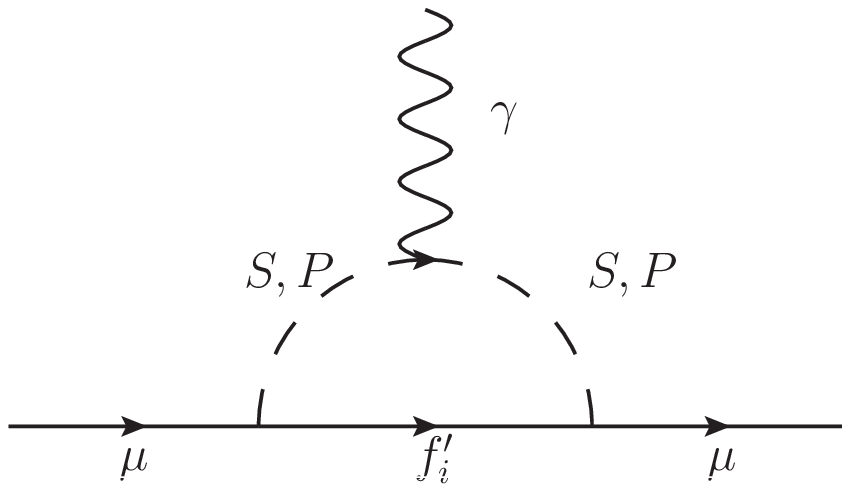}
\label{subfig:CSff}}
\caption{1-loop diagrams contributing to $(g-2)_\mu$.}\label{gen} %Directly replace the out-going muon to electron for $\mu\rightarrow{e} {\gamma}$.}
\end{center}
\end{figure}

Before providing the general methods to reconcile the tension between $(g-2)_\mu$ and the constraint from $\mu\to e\gamma$,
let us first see how general new physics models perturbatively coupled to the SM part can solve the $(g-2)_\mu$ anomaly.
We shall work in a simplified framework in which only the relevant particles and the renormalizable parts of the Lagrangian
related to $(g-2)_\mu$ are given. In this setup, it is enough to only consider the leading one-loop contributions to $\delta a_\mu$.
For simplicity, we confine the spin of the loop particles to be not larger than 1, but we do not restrict their charges as to keep the discussion general.
In this way, the leading one-loop Feynman diagrams can be classified into 4 categories, as explicitly shown in Fig.~\ref{gen},
according to if the boson running in the loop is a (axial-)vector or  (pseudo-)scalar and  the photon is emitted from a fermion or boson.
We calculate these diagrams and present the final expressions for $\delta a_\mu$, which can be compared with those given
in the literature~\cite{Leveille:1977rc,Jegerlehner:2009ry, Lynch:2001zs} with different gauges. In particular,
we also show the analytic formulae in two useful limits: (i) $M_B \gg m_{\mu},\,M_{f^{'}_{i}}$ and (ii) ${M}_B\sim M_{f^{'}_{i}}\gg m_{\mu}$, where $M_B$($M_{f^{'}_{i}}$) denotes the mass of the additional loop boson~(fermion).
However, in a complete model, the total contribution to $(g-2)_\mu$ is usually the summation of the two
or more diagrams in Fig.~\ref{gen}, so the classification here is only for the convenience of the discussion.

\subsection{New Vector Boson}
Besides the new vector boson $X_\mu$, we usually need to introduce some extra fermions $f^{'}_i$ with the internal index $i$. The renormalizable lepton-vector-fermion vertex can be written as follows:
\begin{align}
\label{VecLag}
 	{ \mathcal{L} }_{ int }^{ V }= -\bar { \ell }_\alpha \left\{ ({ C }_{ V  })_{\alpha i}{ \gamma  }^{ \mu  }+({ C }_{ A })_{\alpha i}{ \gamma  }^{ \mu  }{ \gamma  }^{ 5 } \right\} { f }_{ i }^{ ' }{ X }_{ \mu  }+{\rm h.c.}\,,
	 \end{align}
where the subscript $\alpha$ is the charged flavor index and $C_{V(A)}$ the (axial-)vector coupling matrix. 
With this setup, Fig.~\ref{subfig:NVff} and \ref{subfig:CVff} can be calculated as follows.

%\begin{enumerate}
\subsubsection{Photon Emitted From the Internal Fermions (Fig.~\ref{subfig:NVff})}

The contributions to $(g-2)_\mu$ due to the vector (V) and axial-vector (A) couplings are given by~\cite{Jegerlehner:2009ry}
\begin{eqnarray}\label{FV}
{ \delta { a }_{ \mu  }^{ V } }&=&-{ Q }_{ { f }_{ i }^{ ' } }\frac { { |({ C }_{ V })_{\mu i}|^{ 2 } } }{ 4{ \pi  }^{ 2 } } \frac { { m }_{ \mu  }^{ 2 } }{ { M_{V}  }^{ 2 } } {\cal I}^V \nonumber\\
{\cal I}^V &=& \frac{1}{2}\int _{ 0 }^{ 1 }{ dx } \frac { 2x(1-x)(x-2+2\epsilon )+{  \lambda  }^{ 2 }{ (1-\epsilon ) }^{ 2 }{ x }^{ 2 }(1-x+\epsilon ) }{ { (1-x)(1-\lambda ^{ 2 }x) }+({ \epsilon \lambda ) }^{ 2 }x }  \nonumber\\
&=& -\frac{1}{12(1-z)^4}(8-38 z + 39 z^2 -14 z^3 +5 z^4 -18 z^2\ln z) \nonumber\\
 && +\frac{\epsilon}{4(1-z)^3}(4-3z-z^2+6z\ln z) ,\, {\rm if} \, M_V \sim M_{f^\prime_i}\gg m_\mu
\end{eqnarray}
and 
\begin{eqnarray}\label{FA}
\delta { { a }_{ \mu  }^{ A } }&=&-{ Q }_{ { f }_{ i }^{ ' } }\frac { { |({ C }_{ A })_{\mu i}|^{ 2 } } }{ 4{ \pi  }^{ 2 } } \frac { { m }_{ \mu  }^{ 2 } }{ { M_{V}  }^{ 2 } } {\cal I}^{A} \nonumber\\
{\cal I}^A &=& \frac{1}{2}\int _{ 0 }^{ 1 }{ dx } \frac { 2x(1-x)(x-2-2\epsilon )+{ \lambda  }^{ 2 }{ (1+\epsilon ) }^{ 2 }{ x }^{ 2 }(1-x-\epsilon ) }{ { (1-x)(1-\lambda ^{ 2 }x) }+({ \epsilon \lambda ) }^{ 2 }x }  \nonumber\\
&=& -\frac{1}{12(1-z)^4} (8-38 z + 39 z^2 -14 z^3 +5 z^4 -18 z^2\ln z) \nonumber\\
&& -\frac{\epsilon}{4(1-z)^3}(4-3z-z^2+6z\ln z) ,\,{\rm if}\, M_V \sim M_{f^\prime_i}\gg m_\mu
\end{eqnarray}
respectively, where $M_{V}$ is the mass of the vector boson,
$\epsilon ={ { M }_{ { f }_{ i }^{ ' }} }/{ { m }_{ \mu  } }$, $\lambda ={ { m }_{ \mu  } }/{  M_{V}  }$, $z=(\epsilon\lambda)^2=(M_{f^\prime_i}/M_V)^2$ and ${ Q }_{ { f }_{ i }^{ ' } }$ is the electric charge of ${ f }_{ i }^{ ' }$. If we further restrict to the limit of $M_V \gg m_\mu , M_{f^\prime_i}$, the two integrals, ${\cal I}^V$ and ${\cal I}^A$, can be simplified to
\begin{eqnarray}
{\cal I}^V &=& \epsilon -\frac{2}{3}, \nonumber\\
{\cal I}^A &=& -\epsilon-\frac{2}{3}.
\end{eqnarray}
As a simple check of our general formulae, Eqs.~(\ref{FV}) and (\ref{FA}), we can identify $X$ as the $Z$ boson and ${ f }_{ i }^{ ' }$ as the $\mu$ lepton, resulting in
 \begin{align}
 \label{eq:SMZ}
\delta { a }_{ \mu  }^{ Z }=\delta { { a }_{ \mu  }^{ V } }+\delta { { a }_{ \mu  }^{ A } }&=-\frac { { G }_{ F }{ m }_{ \mu  }^{ 2 } }{ 8\sqrt { 2 } { \pi  }^{ 2 } } \frac { 4 }{ 3 } \left( 1+2\sin ^{ 2 }{ { \theta  }_{ W } } -4\sin ^{ 4 }{ { \theta  }_{ W } }  \right)\, ,
 \end{align}
which agrees with the usual SM calculations in the literature~\cite{Beringer:1900zz}. Here, we have used ${ Q }_{ { f'}_{ i } }=-1,~\epsilon=1$, $C^{ 2 }_{ V }= g^{\prime 2 } ( - 1/ 2  +2\sin^2{\theta }_{ W }  )^{ 2 }/4$, $C^{ 2 }_{ A }={ ( { g^\prime}/{ 4 }  )  }^{ 2 }$,  $g'={g_{L}}/{\cos{ { \theta  }_{ W } }}$, and
$M_{W}=M_{Z}\cos{ { \theta  }_{ W } }$. 

Furthermore, since the integral for the purely axial-vector coupling is always negative, the axial-vector contribution from negative charged fermions in the loop is always desconstructive with the SM one. Clearly, there is no hope to add a neutral vector boson with the purely axial-vector coupling  to explain the $(g-2)_\mu$ anomaly.

\subsubsection{ Photon Emitted From A Charged Vector Boson (Fig.~\ref{subfig:CVff})}

In order to differentiate from the previous case, we denote the two general contributions to 
$(g-2)_\mu$ as those from the charged vector (CV) and charged axial-vector (CA) couplings, given by~\cite{Jegerlehner:2009ry}
\begin{eqnarray}\label{eq:CV}
\delta { { a }_{ \mu  }^{ CV } }&=&-{ Q }_{ V }\frac { { |({ C }_{ V })_{\mu i}|^{ 2 } } }{ 4{ \pi  }^{ 2 } } \frac { { m }_{ \mu  }^{ 2 } }{ {  M_{V}  }^{ 2 } }{\cal I}^{CV}\nonumber\\
{\cal I}^{CV}&=& \frac { 1 }{ 2 } \int _{ 0 }^{ 1 }{ dx } \frac { { { 2x }^{ 2 } }(1+x-2\epsilon )-{ \lambda  }^{ 2 }{ (1-\epsilon ) }^{ 2 }x(1-x)(x+\epsilon ) }{ x+({ \epsilon \lambda ) }^{ 2 }(1-x)(1-{ \epsilon  }^{ -2 }x) }  \nonumber\\
&=&\frac{1}{12(1-z)^4}(10-43 z + 78 z^2 -49 z^3 +4 z^4 +18 z^3\ln z) \nonumber\\
 && -\frac{\epsilon}{4(1-z)^3}(4-15z+12z^2- z^3- 6z^2\ln z) ,\, {\rm if} \, M_V \sim M_{f^\prime_i}\gg m_\mu ,
\end{eqnarray}
and
\begin{eqnarray}\label{eq:CA}
\delta { { a }_{ \mu  }^{ CA } }&=&-{ Q }_{ V }\frac { { |({ C }_{ A })_{\mu i}|^{ 2 } } }{ 4{ \pi  }^{ 2 } } \frac { { m }_{ \mu  }^{ 2 } }{ {  M_{V}  }^{ 2 } }{\cal I}^{CA}\nonumber\\
{\cal I}^{CA}
&=&\frac { 1 }{ 2 } \int _{ 0 }^{ 1 }{ dx } \frac { { { 2x }^{ 2 } }(1+x+2\epsilon )-{ \lambda  }^{ 2 }{ (1+\epsilon ) }^{ 2 }x(1-x)(x-\epsilon ) }{ x+({ \epsilon \lambda ) }^{ 2 }(1-x)(1-{ \epsilon  }^{ -2 }x) }   \nonumber\\
&=&\frac{1}{12(1-z)^4}(10-43 z + 78 z^2 -49 z^3 +4 z^4 +18 z^3\ln z) \nonumber\\
 && +\frac{\epsilon}{4(1-z)^3}(4-15z+12z^2- z^3- 6z^2\ln z) ,\,{\rm if}\, M_V \sim M_{f^\prime_i}\gg m_\mu ,
\end{eqnarray}
respectively, where ${ Q }_{ V }$ is the vector boson electric charge. In the limit of $M_V \gg m_\mu , M_{f^\prime_i}$, {\it e.g.}, ${\cal I}^{CV}$ and ${\cal I}^{CA}$ have the following simple form
\begin{eqnarray}
{\cal I}^{CV} &=& -\epsilon +\frac{5}{6}, \nonumber\\
{\cal I}^{CA} &=& \epsilon+\frac{5}{6}.
\end{eqnarray}
We can check Eqs.~(\ref{eq:CV}) and (\ref{eq:CA}) easily by identifying the charged boson to be the W boson with $C^{2}_{V}=C^{2}_{A}=(g_{L}/{ 2\sqrt { 2 }  })^{ 2 }$ and ${ Q }_{ B }=-1$, giving
\begin{align}
\label{eq:SMW}
\delta { a }_{ \mu  }^{ W }=\delta { { a }_{ \mu  }^{ CV } }+\delta { { a }_{ \mu  }^{ CA } }=\frac { { G }_{ F }{ m }_{ \mu  }^{ 2 } }{ 8\sqrt { 2 } { \pi  }^{ 2 } } \frac { 10 }{ 3 } ,
\end{align}	
which is in agreement with the usual SM result at one-loop level~\cite{Beringer:1900zz}.

Note that there are no cross terms of the vector and axial-vector couplings, which are proportional to $C_V C_A^\dagger$ or $C_V^\dagger C_A$ in the $(g-2)_\mu$ contribution, since such terms lead to the muon electric dipole operator, rather than the magnetic dipole one which we are interested in. In addition, our derivation is based on the Lagrangian of Eq.~(\ref{VecLag}), in which the vector boson coupling is decomposed into the vector and axial-vector couplings. However, it is more useful to work on the basis in which the fermion field has definite chirality. The problem is how to use our general formulae Eqs.~(\ref{FV}), (\ref{FA}), (\ref{eq:CV}) and (\ref{eq:CA}) in this chiral basis. We find that if the two vertices involving $X$ are both left-handed or both right-handed, {\emph i.e.}, the internal fermion does not flip its chirality, the expression should be $\delta { { a }_{ \mu  }^{ (C)V } }+\delta { { a }_{ \mu  }^{ (C)A } }$. But, if one vertex is left-handed while the other is right-handed, the internal fermion flips its chirality with the result of $\delta { { a }_{ \mu  }^{ (C)V } }-\delta { { a }_{ \mu  }^{ (C)A } }$.

\subsection{New Scalar/Pseudoscalar Boson}
For new scalar (S) and pseudoscalar (P) bosons, the couplings to the SM leptons $\ell_\alpha$ require to have some additional fermions $f^{'}_{i}$ through the following Yukawa couplings:
\begin{eqnarray}
 	  {  \mathcal{L} }_{ int }^{ S } &=& { Y }_{ \alpha i }\bar { { \ell  }_{ \alpha } } S{ f }_{ i }^{ ' }+{\rm h.c.} ,\label{LagS}\\
 	  { \mathcal{L}  }_{ int }^{ P } &=& { Y }^{'}_{ \alpha i }\bar { { \ell  }_{ \alpha } }i { \gamma  }^{ 5 }P{ f }_{ i }^{ ' }+ {\rm h.c.},\label{LagP}
\end{eqnarray}
where $Y_{\alpha i}$ and ${ Y }^{'}_{ \alpha i }$ are Yukawa coupling matrices for scalars and pseudoscalars, respectively, and $\alpha$ denotes the lepton flavor. The relevant Feynman diagrams are depicted in Figs.~\ref{subfig:NSff} and \ref{subfig:CSff}.

\subsubsection{Photon emitted from internal fermion (Fig. \ref{subfig:NSff})}\label{SecFermion}

In order for this diagram to contribute to $(g-2)_\mu$, the internal fermions should have charge $Q_{f_i^{'}}$, no matter if the (pseudo)scalar is charged or not. Moreover, it is convenient to further divide the $\delta a_\mu$ contribution into two parts: scalar (S) and pseudoscalar (P) ones, since they lead to the different expressions.

\begin{figure}[ht]
\centering
 \subfigure[short for lof][Scalar]{
   \includegraphics[width=0.45\linewidth]{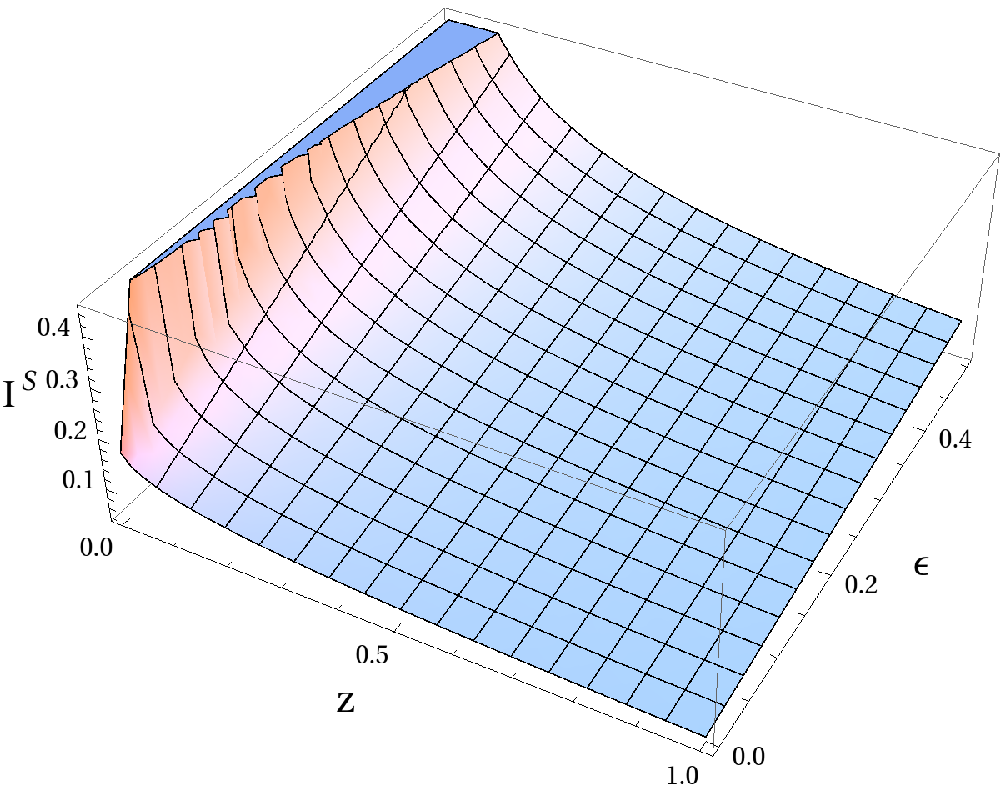}
   \label{subfig:NS}
 }
 \subfigure[short for lof][Pseudoscalar]{
   \includegraphics[width=0.45\linewidth]{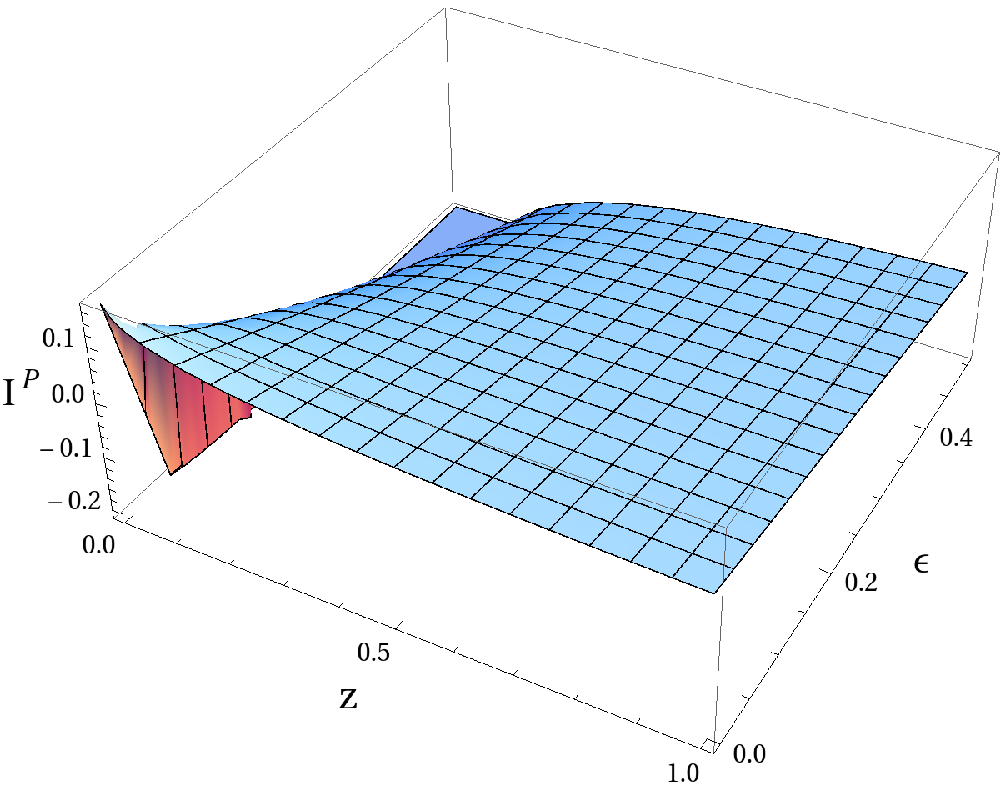}
   \label{subfig:NP}
}
\caption[short for lof]{The integration parts of (a) $\delta { { a }_{ \mu  }^{ S } }$ and (b) $\delta { { a }_{ \mu  }^{ P } }$ in the limits of $M_{S} \gg m_{\mu},M_{f^{'}_{i}}$.}
\label{fig:fig1}
\end{figure}

\noindent {\bf  Scalar}:
The general contribution to $(g-2)_\mu$ from a scalar (S) coupling in Eq.~(\ref{LagS}) is	
\begin{eqnarray}\label{eq:NS}
\delta { { a }_{ \mu  }^{ S } }&=&-{ Q }_{ { f }_{ i }^{ ' } }\frac { { |{ Y }_{ \mu i }|^{ 2 } } }{ 4{ \pi  }^{ 2 } } \frac { { m }_{ \mu  }^{ 2 } }{ { M_{S}  }^{ 2 } }{\cal I}^{S}\nonumber\\
{\cal I}^{S}&=& \frac { 1 }{ 2 }\int _{ 0 }^{ 1 }{ dx } \frac { { x }^{ 2 }(1-x+\epsilon ) }{ { (1-x)(1-\lambda ^{ 2 }x) }+({ \epsilon \lambda ) }^{ 2 }x }  \nonumber\\
&=&\frac{1}{12(1-z)^4}(2+ {3} z- 6 z^2 + z^3 +6 z\ln z) \nonumber\\
 && -\frac{\epsilon}{4(1-z)^3}(3-4z+z^2+ 2z\ln z),\, {\rm if} \, M_S \sim M_{f^\prime_i}\gg m_\mu, \label{eq:NSc}
\end{eqnarray}
where $M_{S}$ is the scalar mass, $\epsilon ={ { M }_{ { f }_{ i }^{ ' }} }/{ { m }_{ \mu  } }$, $\lambda ={ { m }_{ \mu  } }/{  M_{S}  }$, $z=(\epsilon\lambda)^2=(M_{f^\prime_i}/M_S)^2$, and ${ Q }_{ { f }_{ i }^{ ' } }$ is the charge of ${ f }_{ i }^{ ' }$ as before. In the limit of $M_S \gg m_\mu , M_{f^\prime_i}$, we can further simplify the integral ${\cal I}^{S}$ to
\begin{eqnarray}
{\cal I}^{S} &=& -\epsilon \ln{ \left( \frac { { M }_{ { f }_{ i }^{ ' } } }{ M_{S}  }  \right)  }-\frac { 3 }{ 4 } \epsilon +\frac { 1 }{ 6 }.
\end{eqnarray}
Note that the integral $ \mathcal{I}^{S}$ in Eq.~(\ref{eq:NS}) does not have the definite sign in the parameter region we are interested in. As a result, it is more useful to plot $ \mathcal{I}^{S}$ against $\epsilon$ and $z$ on the 3D diagram in Fig. \ref{subfig:NS}.
We find that in the limits $M_{S}\gg m_{\mu},M_{f^{'}_{i}}$ and ${M_{S}}\sim M_{f^{'}_{i}}\gg m_{\mu}$, the value of $I^{S}$ is
 always positive. Thus, the sign of $\delta a_\mu^S$ only depends on $Q_{f^{'}}$. For example, if we take the fermion charge to be $Q_{f^{'}}=-1$ and the scalar to be neutral, the new contribution to the muon $g-2$ is constructive with the SM one. We can obtain $\Delta a_\mu = 287\times 10^{-11}$ by choosing ${ Y }_{ \mu i }\sim 10^{-2}$ and $ M_S \sim { M }_{ { f }_{ i }^{ ' } } \sim 1 $~TeV, promising to be probed by the next run of the LHC experiments.

\noindent {\bf  Pseudoscalar}:
The pseudoscalar (P) coupling in Eq.~(\ref{LagP}) can give the following general $(g-2)_\mu$ contribution,
\begin{eqnarray}\label{eq:NP}
\delta { { a }_{ \mu  }^{ P } }&=&-{ Q }_{ { f }_{ i }^{ ' } }\frac { { |{ Y }_{ \mu i }|^{ 2 } } }{ 4{ \pi  }^{ 2 } } \frac { { m }_{ \mu  }^{ 2 } }{ { M_{P}  }^{ 2 } }{\cal I}^{P}\nonumber\\
{\cal I}^{P}&=&\frac{1}{2}\int _{ 0 }^{ 1 }{ dx } \frac { { x }^{ 2 }(1-x-\epsilon ) }{ { (1-x)(1-\lambda ^{ 2 }x) }+({ \epsilon \lambda ) }^{ 2 }x } \nonumber\\
&=&\frac{1}{12(1-z)^4}(2+3 z- 6 z^2 + z^3 +6 z\ln z) \nonumber\\
 && +\frac{\epsilon}{4(1-z)^3}(3-4z+z^2+ 2z\ln z),\, {\rm if} \, M_P \sim M_{f^\prime_i}\gg m_\mu, \label{eq:NPc}
\end{eqnarray}
where $M_{P}$ denotes the pseudoscalar mass, $\epsilon ={ { M }_{ { f }_{ i }^{ ' }} }/{ { m }_{ \mu  } }$, $\lambda ={ { m }_{ \mu  } }/{  M_{P}  }$, and $z=(\epsilon\lambda)^2=(M_{f^\prime_i}/M_P)^2$. If we further restrict to the limit of $M_P \gg m_\mu , M_{f^\prime_i}$, the integral ${\cal I}^{P}$ can be reduced  to
\begin{eqnarray}
{\cal I}^{P} &=&  \epsilon \ln{ \left( \frac { { M }_{ { f }_{ i }^{ ' } } }{ M_{P}  }  \right)  }+\frac { 3 }{ 4 } \epsilon +\frac { 1 }{ 6 }.
\end{eqnarray}
As for the scalar case, we also plot $\mathcal{I}^{P}$ in Eq.~(\ref{eq:NP}) against $\epsilon$ and $z$ on the 3D diagram in Fig. \ref{subfig:NP}. It is useful to note that most of the parameter space is negative except for the $\epsilon\equiv M_{f^{'}_{i}} /m_{\mu}\lesssim 0.2$ region.

As an example, if ${ Q }_{ { f }_{ i }^{ ' } }=-1$ and the pseudoscalar is neutral, the sign of its contribution to the muon $g-2$ is positive only when $\epsilon \ll 1$, {\it i.e.}, $M_{f^{'}_{i}} \ll m_{\mu}$, as seen from Fig.~\ref{subfig:NP}. However, if there exists such a light charged fermion,  it should have already been observed at the colliders, such as the LEP. Therefore, a model with a neutral pseudoscalar and a light charged fermion favored by the current $(g-2)_\mu$ data is already ruled out.

\subsubsection{Photon emitted from charged scalar or pseudoscalar particles (Fig.~\ref{subfig:CSff})}\label{SecScalar}

The calculation of the Feynman diagram in Fig.~\ref{subfig:CSff} leads to
the  contribution to the muon AMM from loops with a charged scalar (CS) or pseudoscalar~(CP).

\noindent {\bf Scalar}:
When the photon is emitted from a charged scalar, the general contribution to the muon $g-2$ in Fig.~\ref{subfig:CSff} is
\begin{eqnarray}\label{eq:CS}
\delta { { a }_{ \mu  }^{ CS } }&=&-{ Q }_{ S }\frac { { |{ Y }_{ \mu i }|^{ 2 } } }{ 4{ \pi  }^{ 2 } } \frac { { m }_{ \mu  }^{ 2 } }{ { M_{S}  }^{ 2 } }{\cal I}^{CS} \nonumber\\
{\cal I}^{CS}&=&\frac{1}{2}\int _{ 0 }^{ 1 }{ dx } \frac { { x }(x-1)(x+\epsilon ) }{ x+({ \epsilon \lambda ) }^{ 2 }(1-x)(1-{ \epsilon  }^{ -2 }x) }\\
&=&\frac{-1}{12(1-z)^4}(1-6 z+ 3 z^2 +2 z^3 -6 z^2 \ln z) \nonumber\\
 && -\frac{\epsilon}{4(1-z)^3}(1-z^2+ 2z\ln z),\, {\rm if} \, M_S \sim M_{f^\prime_i}\gg m_\mu , \label{eq:CSc}
\end{eqnarray}
where ${ Q }_{ S }$ is the scalar charge, $\epsilon ={ { M }_{ { f }_{ i }^{ ' }} }/{ { m }_{ \mu  } }$, $\lambda ={ { m }_{ \mu  } }/{  M_{S}  }$ and $z=(\epsilon\lambda)^2 = (M_{f^\prime_i}/M_S)^2$ as before. When $M_S \gg m_\mu , M_{f^\prime_i}$, ${\cal I}^{CS}$ can be simplified to
\begin{eqnarray}
{\cal I}^{CS} &=&  -\frac { 1 }{ 4 } \epsilon -\frac { 1 }{ 12 }.
\end{eqnarray}
The integral in Eq.~(\ref{eq:CS}) is always negative. Consequently, if ${ Q }_{ S }$ is minus, $\delta { { a }_{ \mu  }^{ CS } }$ will depress $a_{\mu}$. The simple application of this result is that when the loop in Fig.~\ref{subfig:CSff} is enclosed by a scalar with $Q_S =-1$ and a neutral fermion, the contribution can never reduce the tension between the SM prediction and data for the muon $g-2$.

\noindent {\bf Pseudoscalar}:
When the scalar in Fig.~\ref{subfig:CSff} is replaced by the charged pseudoscalar, the expression for the $(g-2)_\mu$ contribution becomes,
\begin{eqnarray}\label{eq:CP}
\delta { { a }_{ \mu  }^{ CP } } &=& -{ Q }_{ P }\frac { { |{ Y }^{'}_{ \mu i }|^{ 2 } } }{ 4{ \pi  }^{ 2 } } \frac { { m }_{ \mu  }^{ 2 } }{ { M_{P}  }^{ 2 } }{\cal I}^{CP}\nonumber\\
{\cal I}^{CP}&=&\frac{1}{2}\int _{ 0 }^{ 1 }{ dx } \frac { { x }(x-1)(x-\epsilon ) }{ x+({ \epsilon \lambda ) }^{ 2 }(1-x)(1-{ \epsilon  }^{ -2 }x) } \\
&=&\frac{-1}{12(1-z)^4}(1-6 z+ 3 z^2 +2 z^3 -6 z^2 \ln z) \nonumber\\
 && +\frac{\epsilon}{4(1-z)^3}(1-z^2+ 2z\ln z),\, {\rm if} \, M_P \sim M_{f^\prime_i}\gg m_\mu , \label{eq:CPc}
\end{eqnarray}
where ${ Q }_{ P }$ is the charge of the pseudoscalar, $\epsilon ={ { M }_{ { f }_{ i }^{ ' }} }/{ { m }_{ \mu  } }$, $\lambda ={ { m }_{ \mu  } }/{  M_{P}  }$ and $z= (\epsilon\lambda)^2 =(M_{f^\prime_i}/M_P)^2$. The integral ${\cal I}^{CP}$ in the limit $M_P \gg m_\mu , M_{f^\prime_i}$ becomes
\begin{eqnarray}
{\cal I}^{CP} &=&  \frac { 1 }{ 4 } \epsilon -\frac { 1 }{ 12 }.
\end{eqnarray}
For a rough estimation, with a pseudoscalar with $Q_P=-1$ and a neutral fermion in the loop, the Feynman diagram in Fig.~\ref{subfig:CSff} can solve the muon $g-2$ anomaly with $Y^{'}_{\mu i}\sim10^{-2}$ and $M_P\sim { M }_{ { f }_{ i }^{ ' } }\sim10$~TeV.

%\end{enumerate}	

Similar to the (axial-)vector case in the previous subsection, the cross terms of the scalar and pseudoscalar couplings, which are proportional to $Y Y^{'\dagger}$ or $Y^{\dagger} Y^{'}$, do not contribute to the muon $g-2$. Rather, they give rise to the muon electric dipole moment. Moreover, when the scalar/pseudoscalar couplings in Eqs.~(\ref{LagS}) and (\ref{LagP}) are decomposed into the basis in which the chiralities of the fermion fields are well defined, we can easily obtain the $(g-2)_\mu$ formulae from Eqs.~(\ref{eq:NS}), (\ref{eq:NP}), (\ref{eq:CS}) and (\ref{eq:CP}). Concretely, when the two vertices on the fermion line are both left- and right-handed, the result would be  $\delta { { a }_{ \mu  }^{ (C)S } }-\delta { { a }_{ \mu  }^{ (C)P } }$. 
Note that the internal fermion has to flip its chirality in this case so that the final expression should be proportional to the internal fermion mass. However, for the case that the two couplings have different chiralities, the result is $\delta { { a }_{ \mu  }^{ (C)S } }+\delta { { a }_{ \mu  }^{ (C)P } }$.

%{\color{red} 
It is interesting to note that our general formalism can be applied to the general supersymmetric extension of the SM. 
One particular example is the minimal supersymmetric Standard Model (MSSM), in which neutralinos, charginos and various sleptons provide the additional contributions to $(g-2)_\mu$ and $\mu \to e\gamma$. Especially, one-loop diagrams related to charginos and neutralinos correspond to our case in Sec.~\ref{SecFermion} and \ref{SecScalar}, so the formulae in these two subsections can be used directly. The detailed discussions of
 $(g-2)_{\mu}$ and its correlation to $\mu \to e\gamma$ are summarized in Ref.~\cite{Stockinger:2006zn}.
 %}
 %,Moroi:1995yh, Chacko:2001xd, Kersten:2014xaa}. }

\section{Some General Solutions to the Tension Between the $(g-2)_\mu$ anomaly and the $\mu\rightarrow{e}{\gamma}$ Constraint}
Now, we come back to the question how to reconcile the tension between the muon $g-2$ anomaly and the $\mu \to e \gamma$ constraint in a generic new physics model. Here, we provide two generic methods mentioned in Sec.~\ref{intro}.

\subsection{GIM Mechanism}
As is well known in the SM, there is no flavor-changing neutral current (FCNC) at tree level due to the famous GIM mechanism~\cite{Glashow:1970gm}. Even at loop level, the GIM mechanism also suppresses the FCNC greatly. Of our present interest is the SM contributions to $(g-2)_\mu$ and $\mu \to e\gamma$. In the SM with the non-zero neutrino masses, the one for $\mu \to e\gamma$ can be obtained by the loops running a $W$-boson with different flavors of neutrinos. But because of the GIM mechanism, the unitarity of the PMNS matrix
results in the leading terms to be canceled. The terms left are at least proportional to the square of the neutrino masses. Since the neutrino masses are negligibly small, it is no hope to observe the $\mu \to e \gamma$ rate in the current experiments.
However, $(g-2)_\mu$ does not suffer such a huge suppression, leaving us measurable signals. We hope that the GIM mechanism may also happen in the new physics sector beyond the SM.

To be specific, let us consider a model in which a vector $X$ and some fermions $f_i$ with masses $m_i$ are introduced with the following chiral couplings:
\begin{equation}\label{LagGIM}
{\cal L} = g \bar{\ell}_\alpha U_{\alpha i} \gamma^\mu P_L f_i X_\mu + {\rm h.c.},
\end{equation}
where  $\ell_\alpha$ the different charged leptons in the SM with the flavor index $\alpha$. We have extracted the overall coupling constant $g$ to make the mixing matrix $U_{\alpha i}$ to satisfy the following orthogonal normalization conditions,
\begin{align}\label{Unitary}
\sum _{ i }{ { U }_{ \alpha i }{ U }_{\beta i }^{ \ast  } } &= \left\{\begin{array}{ll}
1, & \mbox{if $\alpha=\beta.$} \\
0, & \mbox{if $\alpha\neq\beta.$} \\
\end{array} \right.   ,
\end{align}
where $\alpha$ and $\beta$ are the flavor indices of charged leptons ($e$, $\mu$, $\tau$). Here, the mixing matrix $U_{\alpha i}$ is not necessarily unitary but can be extended to a rank-three $3\times n$ one, where $n$($\geq 3$) is the number of the internal fermions. Note that the Lagrangian in Eq.~(\ref{LagGIM}) is very similar to the SM W-boson couplings with the W-boson replaced by $X_\mu$ and the neutrinos by $f_i$. However, to make the discussion more general, we take the charge of the vector boson $X_\mu$ to be $Q_X$ so that the fermion charge should be $Q_i=-1-Q_X$. We also assume a mass hierarchy $m_X \gg m_i \gg m_\ell$ ($\ell = e,~\mu, ~ \tau$) for simplicity.

With the Lagrangian in Eq.~(\ref{LagGIM}), the leading-order one-loop Feynman diagrams relevant to $\mu \to e \gamma$ are essentially the same as the first two diagrams in Fig.~2 except for the outgoing $\mu$ replaced by $e$, resulting in the following effective operator,
\begin{eqnarray}\label{EffGIM}
\delta {\cal L}^{ \mu \rightarrow { e }{ \gamma  } }_{ eff } =  \bar { e } { \sigma  }_{ \mu \nu  } P_R \mu { F }^{ \mu \nu  } (g^2 m_\mu) \sum _{ i }{ { U }_{ \mu i }^{ \ast  }{ U }_{ ei } } \bar{C}(m_X, m_i, m_e, m_\mu)
\end{eqnarray}
where
\begin{equation}
\bar{C} (m_X, m_i, m_e, m_\mu) = \left\{ Q_X {\cal I}_1(m_X, m_i, m_e, m_\mu) + Q_i {\cal I}_2(m_X, m_i, m_e, m_\mu)  \right\},
\end{equation}
and the integrals ${\cal I}_{1,2}$ are the separated expressions for the two Feynman diagrams in Fig.~2. The explicit expressions of ${\cal I}_{1,2}$ are not given since they are not very crucial for our discussion of the general GIM mechanism. What we should focus is the track of the coupling factors, especially the mixing matrix elements $U_{\alpha i}$,  which can be easily read out from the Feynman diagrams. Note that the two external charged leptons in the effective vertex $\delta {\cal L}^{ \mu \rightarrow { e }{ \gamma  } }_{ eff }$ have different chiralities, while the interaction in Eq.~(\ref{LagGIM}) only involves left-handed fermions. Therefore, an extra lepton mass insertion is needed to obtain $\delta {\cal L}^{ \mu \rightarrow { e }{ \gamma  } }_{ eff }$ by flipping the chirality of either lepton, producing two terms proportional to $m_e$ and $m_\mu$. However, since $m_e\ll m_\mu$, only the latter term is kept, which is the origin of the factor $m_\mu$ and the right-handed projection operator $P_R$ in Eq.~(\ref{EffGIM}). Moreover, the mass dimensions of the integrals ${\cal I}_{1,2}$ are $-2$, which can be made dimensionless by extracting the inverse of the largest mass scale $m_X$ squared. In this way, the total integral $\bar{C}$ can be written in the following form:
\begin{eqnarray}\label{Expansion}
\bar{C} (m_X, m_i, m_e, m_\mu) &=& \frac{1}{m_X^2} C(m_i/m_X, m_e/m_X, m_\mu /m_X)\nonumber\\
&=& \frac{1}{m_X^2} (C_0+ C_2 \frac{m_i^2}{m_X^2} + ...),
\end{eqnarray}
where $C_{2i}$ are the coefficients for the power expansion of $C(m_i/m_X, m_e/m_X, m_\mu /m_X)$ in terms of ${m_i}/{m_X}$, and are usually expected as the ${\cal O}$(1) function of $m_{\mu(e)}/m_X$ . Since only the squares of various particle masses appear in the integral $\bar{C}$, the expansion only has even powers of ${m_i}/{m_X}$.

By putting Eq.~(\ref{Expansion}) back into Eq.~(\ref{EffGIM}), the effective operator for $\mu \to e\gamma$ has the form:
\begin{eqnarray}
\delta {\cal L}^{ \mu \rightarrow { e }{ \gamma  } }_{ eff } =  \bar { e } { \sigma  }_{ \mu \nu  } P_R\mu { F }^{ \mu \nu  } \left(\frac{g^2 m_\mu}{m_X^2}\right) \sum _{ i }{ { U }_{ \mu i }^{ \ast  }{ U }_{ ei } } (C_0+ C_2 \frac{m_i^2}{m_X^2}+...).
\end{eqnarray}
Note that the leading term vanishes due to the orthogonal condition in Eq.~(\ref{Unitary}). We are, then, left with the second-order term
\begin{equation}
\delta {\cal L}^{ \mu \rightarrow { e }{ \gamma  } }_{ eff } =  \bar { e } { \sigma  }_{ \mu \nu  } P_R \mu { F }^{ \mu \nu  } \left(\frac{g^2 m_\mu}{m_X^2}\right) \sum _{ i }{ { U }_{ \mu i }^{ \ast  }{ U }_{ ei } } \left\{C_2 \frac{m_i^2}{m_X^2} + {\cal O}\left( \frac{m_i^4}{m_X^4} \right)\right\}\, ,
\end{equation}
since, in general, we have
\begin{align}
\label{eq:unitary}
    \sum _{ i }{ { U }_{ \mu i }^{ \ast  }{ U }_{ ei } }& { m }^{2}_{ i } \neq 0,
\end{align}
for the different internal fermion masses. By considering the hierarchy $m_X\gg m_i \gg m_\ell$,
we see a large suppression at least in order of $m_i^2/m_X^2$. In contrast, for the flavor conserving muon $g-2$ correction, there is no such suppression due to the normalization condition for the outgoing and incoming leptons with the same flavor. Typically, we expect that the contribution to $(g-2)_\mu$ should be of order:
\begin{eqnarray}
\delta {\cal L}^{a_\mu}_{ eff } =  \bar { \mu } { \sigma  }_{ \mu \nu  } \mu { F }^{ \mu \nu  }
\left(\frac{g^2 m_\mu}{m_X^2}\right) \left(C_0^\prime + {\cal O}\left(\frac{m_i^2}{m_X^2}\right)\right)\,,
\end{eqnarray}
with some ${\cal O}$(1) coefficient $C_0^\prime$. Thus, the LFV process of $\mu\to e\gamma$ is naturally suppressed greatly compared with $\delta a_\mu$.

Note that the GIM mechanism is also applicable to the case when the vector interaction is purely right-handed, rather than left-handed as discussed above. However, if the couplings of the new vector $X_\mu$ to the leptons and the internal fermions have both left-handed and right-handed parts, the GIM mechanism generically breaks down. One reason is that the couplings of different chiralities involve different mixing matrices. For example, let us denote the left-handed vertex mixing matrix by $U$, while the right-handed by $V$. Obviously, the products of two matrices, such as $U^\dagger V$ and $V^\dagger U$,  do not necessarily obey the normalized orthogonal conditions in Eq.~(\ref{Unitary}). Furthermore, when the two vertices in the diagrams shown in Figs.~2(a) and 2(b) have different chiralities, there should be an additional internal fermion mass $m_i$ insertion for flipping its chirality, rather than the much smaller lepton masses. As a result, the leading-order new-physics corrections to $(g-2)_\mu$ and $\mu \to e \gamma$ should be of the same magnitude. Consequently, we can encode this argument in terms of the following effective operators
\begin{eqnarray}
\delta {\cal L}^{a_\mu}_{ eff } &=& \left(\frac{g_L g_R}{m_X^2}\right) \sum_i \bar{\mu} \sigma_{\mu \nu} (U^\ast_{\mu i} V_{\mu i} P_L + V^\ast_{\mu i} U_{\mu i} P_R) \mu F^{\mu \nu} m_i \left\{ C_0^\prime +{\cal O}\left( \frac{m_i^2}{m_X^2} \right) \right\}\, , \nonumber\\
\delta {\cal L}^{ \mu \rightarrow { e }{ \gamma  } }_{ eff } &=& \left(\frac{g_L g_R}{m_X^2}\right) \sum_i \bar{e} \sigma_{\mu \nu} (U^\ast_{\mu i} V_{e i} P_L + V^\ast_{\mu i} U_{e i} P_R) \mu F^{\mu \nu} m_i \left\{ C_0 +{\cal O}\left( \frac{m_i^2}{m_X^2} \right) \right\}\, ,
\end{eqnarray}
where $g_{L,R}$ are left- and right-handed couplings for the vector boson $X_\mu$. In general, the two effective operators have the same order, except for the special case in which all the masses of the internal fermions are the same and $V = U$.

We have discussed the GIM mechanism in a general theory with massive vector bosons. One related question is whether the GIM mechanism still works in a (pseudo)scalar theory. Unfortunately, the Yukawa couplings $Y^{(')}_{\alpha i}$ in Eqs.~(\ref{LagS}) and (\ref{LagP}) are general complex matrices, and do not have the orthogonal properties as in Eq.~(\ref{Unitary}). Thus, the GIM mechanism cannot be applied to an ordinary model with some new (pseudo)scalars.

\subsection{Non-Universal Couplings}
Another possible method to give a large enough $\delta a_{\mu}$ without exceeding the $\mu\rightarrow{e}{\gamma}$
bound is based on the observation that these two processes involve different coupling constants in a new physics model. From the diagrams in Fig.~\ref{gen}, we find that the muon $g-2$ contributions are proportional to either
$|(C_{V(A)})_{\mu i}|^2$ for the (axial-)vectorial couplings in Eq.~(\ref{VecLag}) or
 $|Y^{(')}_{\mu i}|^2$ for the Yukawa couplings in Eqs.~(\ref{LagS}) and (\ref{LagP}), while for the $\mu \to e \gamma$ process, the amplitudes should be proportional to $(C_{V(A)})_{\mu i}^\ast (C_{V(A)})_{e i}$ or $Y^{(')\ast}_{\mu i} Y^{(')}_{e i}$ correspondingly. Since the couplings $(C_{V(A)})_{\mu(e) i}$ and $Y^{(')}_{\mu(e) i}$ are generically free parameters, we have the freedom to choose the coupling matrices $C_{V(A)}$ and $Y^{(')}$ such that the combinations $(C_{V(A)})_{\mu i}^\ast (C_{V(A)})_{e i}$ and $Y^{(')\ast}_{\mu i} Y^{(')}_{e i}$ are always smaller than $|(C_{V(A)})_{\mu i}|^2$ and $|Y^{(')}_{\mu i}|^2$ by at least 5 orders, as indicated by the effective operator analysis in Sec.~\ref{SecEFT}. In this way, we have the possibility to suppress $\mu \to e \gamma$ to the allowed magnitude while still giving a large enough $(g-2)_\mu$ correction to solve the anomaly. Although such a choice of the coupling constants has a little fine-tuning,
it is still acceptable if we consider the same-level hierarchy between the top-quark and the electron Yukawa couplings. In the following, we would like to use a leptoquark model introduced in Ref.~\cite{Arnold:2013cva} as a simple UV-complete theory to exemplify the application of this method.

\subsubsection{$\delta a_\mu$ and $\mu\to e\gamma$ in a scalar leptoquark model}

In Ref.~\cite{Arnold:2013cva}, only one extra scalar leptoquark $X$ is introduced for phenomenological reasons.
In order to eliminate the dangerous tree-level proton decay via $X$, one can find that only two choices of
the leptoquarks are allowed, whose quantum numbers under the SM groups of $SU_C\times SU(2)_L\times U(1)_Y$
are ({\bf 3}, {\bf 2}, 7/3) and ({\bf 3}, {\bf 2}, 1/3), respectively. In the present paper, we will concentrate on the former case and compute its muon $g-2$ contributions from the leptoquark, which was not discussed in the original paper. For other aspects of the model, readers are recommended to refer to Ref.~\cite{Dorsner:2013tla}. The relevant Lagrangian for the leptoquark couplings is given by
\begin{eqnarray}\label{LagLQ}
   \mathcal{L}^{LQ}=-{ \lambda  }_{ u }^{ ij }\overline { { u }_{ R }^{ i } } { X }^{ T }\epsilon { L }_{ L }^{ j }-{ \lambda  }_{ e }^{ ij }\overline { { e }_{ R }^{ i } } { X }^{ \dagger  } { Q }_{ L }^{ j }+{\rm h.c.} \, ,
\end{eqnarray}
with
\begin{align}
\label{}
   X=\begin{pmatrix} { X }_{ 1 } \\ { X }_{ 2 } \end{pmatrix},\quad
   L_{L}=\left(\begin{array}{c} { \nu  }_{ L }\\ { e  }_{ L }  \end{array}\right),\quad Q_{L}=\left(\begin{array}{c} { u  }_{ L }\\ { d  }_{ L }  \end{array}\right),
\end{align}
where $\lambda_{u,e}^{ij}$ denote the Yukawa couplings related to the right-handed $u$-type quarks and the right-handed $e$-type leptons, respectively, and $\epsilon$ is the usual antisymmetric tensor for the $SU(2)_L$ gauge group.

\begin{figure}[ht] % h:here; t:top; b:bottom; p:page; default:ht
\centering
{
   \includegraphics[width=0.6\linewidth]{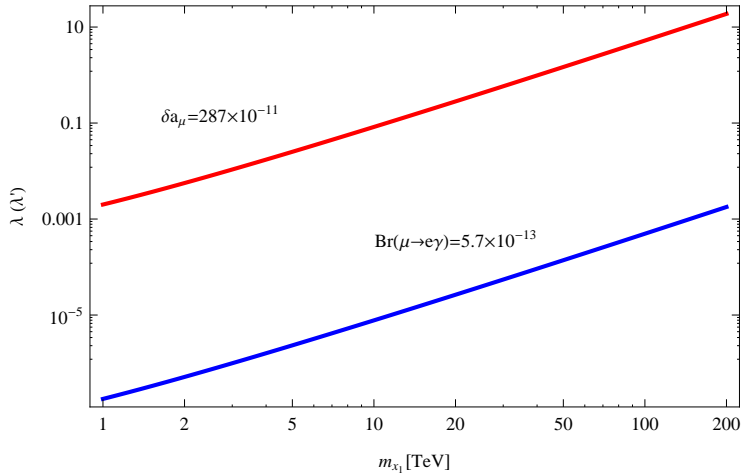}
}
\caption[short for lof] {Contours of $\delta { a }_{ \mu  }=287\times10^{-11}$ (upper curve) and ${\cal B}(\mu \to e\gamma)=5.7\times 10^{-13}$ (lower curve) in the plane of the couplings $\lambda^{(\prime)}$ and the leptoquark mass $m_{X_{1}}$, where $\lambda^{(\prime)}$ is the coupling related to $\mu\to e\gamma$  ($(g-2)_\mu$).}
\label{fig:leptoquarkNew}
\end{figure}

Due to the couplings in Eq.~(\ref{LagLQ}), various one-loop diagrams enclosed by the two leptoquark components and different quark flavors can contribute to $\delta a_\mu$ and $\mu \to e \gamma$. However, by assuming that the leptoquark is much heavier than any SM quarks and leptons, {\it i.e.}, $m_{\mu},m_{Q}\ll m_{X}$, we find that both amplitudes for the muon $g-2$ and $\mu \to e \gamma$ are proportional to the quark masses in the loop~\cite{Arnold:2013cva,Benbrik:2008si}. Thus, the dominant contributions to both phenomena should come from the top-$X_1$ loops. Consequently, the $\mu\rightarrow{e}{\gamma}$ decay rate in this model is given by
\begin{equation}\label{MuE}
    \Gamma (\mu\rightarrow{e}{\gamma})=\frac { { e }^{ 2 }{ \lambda  }^{ 2 }{ m }_{ t }^{ 2 }{ m }_{ \mu  }^{ 3 } }{ 2048{ \pi  }^{ 5 }{ m }_{ { X }_{ 1 } }^{ 4 } } f^{ 2 }\left( \frac { { m }_{ t }^{ 2 } }{ { m }_{ { X }_{ 1 } }^{ 2 } }  \right)  ,
\end{equation}
with
\begin{eqnarray}
f\left( x \right) &=&\frac { 1-{ x }^{ 2 }+2x\log { x }  }{ 2{ \left( 1-x \right)  }^{ 3 } } +\frac { 2 }{ 3 } \left( \frac { 1-x+\log { x }  }{ { \left( 1-x \right)  }^{ 2 } }  \right) ,\label{f-function} \\
 \lambda &\equiv & \sqrt { { \frac { 1 }{ 2 } \left| { \tilde { \lambda  }  }_{ e }^{ 13 }{ \tilde { \lambda  }  }_{ u }^{ 32 } \right|  }^{ 2 }+{ \frac { 1 }{ 2 } \left| { \tilde { \lambda  }  }_{ u }^{ 31 }{ \tilde { \lambda  }  }_{ e }^{ 23 } \right|  }^{ 2 } },
 \end{eqnarray}
where,
\begin{align}\label{}
       { \tilde { \lambda  }  }_{ u }={ U(u,R) }^{ \dagger  }{ \lambda  }_{ u }U(e,L),\quad { \tilde { \lambda  }  }_{ e }={ U(e,R) }^{ \dagger  }{ \lambda  }_{ e }U(u,L) \, ,
\end{align}
and $U(f,L(R))$ denotes the mixing matrix that brings the left-handed (right-handed) fermions from the flavor to the mass eigenstates. On the other hand, from Eqs. \eqref{eq:NSc}, \eqref{eq:NPc}, \eqref{eq:CSc} and \eqref{eq:CPc}, the dominant contribution to $(g-2)_\mu$ with internal a top quark is~\cite{Cheung:2001ip}
\begin{eqnarray}
\label{eq:letoquarkg-2}
\delta { a }_{ \mu  } &=& - \frac{N_c}{4\pi^2}\frac { { m }_{ \mu  }^{ 2 } }{ { m }_{  X_1  }^{ 2 } }  \Big\{ \frac{1}{4} ({|{\tilde{\lambda}}_{e}^{23}|}^{2}+{|{\tilde{\lambda}}_{u}^{32}|^2 } )  \left[ { Q }_{ t }\left( {\cal I}^{S}+ {\cal I}^{P} \right)+{ Q }_{  X_1 } \left( {\cal I}^{CS}+ {\cal I}^{CP} \right) \right] ) \nonumber\\
&& + \frac{1}{2}{\rm Re}\left[ { \tilde { \lambda  }  }_{ e }^{ 23 }{ \tilde { \lambda  }  }_{ u }^{ 32  }\right] \left[ { Q }_{ t }\left( {\cal I}^{S}- {\cal I}^{P} \right)+{ Q }_{  X_1 } \left( {\cal I}^{CS}- {\cal I}^{CP} \right) \right]   \Big\} \nonumber\\
&\approx & \frac{3\lambda^\prime}{4 \pi^2} \frac{m_\mu m_t}{m_{X_1}^2}  \left(\frac{2}{3}  \ln  \frac{m_{X_1}}{m_t} - \frac{1}{12}   \right),
\end{eqnarray}
where
\begin{align}
\label{eq: lambda'}
\lambda '&\equiv  -{\rm Re} \left[ { \tilde { \lambda  }  }_{ e }^{ 23 }{ \tilde { \lambda  }  }_{ u }^{ 32  }\right]  ,
\end{align}
$N_c =3$ is the number of colors, and $Q_{{ X }_{ 1 } (t)}=-5/3 (2/3)$ is the charge of $X_{1}$($t$). We also have taken the limit $M_{X}\gg m_{f'}, m_{\mu}$ in Eq.~(\ref{eq:letoquarkg-2}) to simplify the final result. With the formulae in Eqs.~(\ref{MuE}) and (\ref{eq:letoquarkg-2}), we can plot the contour of $\delta { a }_{ \mu  }=287\times10^{-11}$ and the boundary of the allowed region ${\cal B}(\mu \to e \gamma) < 5.7\times 10^{-13}$ in the $m_{X_1}$-$\lambda^{(\prime)}$ plane, as shown in Fig.~\ref{fig:leptoquarkNew}.
Note that the couplings $\lambda$ and $\lambda^\prime$ have  totally different dependences on the more fundamental Yukawa couplings $\lambda_{u(e)}$, so that they actually have no direct relations. We can tune these Yukawa couplings to make $\lambda^\prime$ positive and larger than $\lambda$ by at least four orders of magnitude so as to fit the $a_\mu$ deviation and suppress $\mu \to e \gamma$ to the allowed order simultaneously, which represents the essence of the method of the non-universal couplings.

Note that as the process $\mu \to e\gamma$ does not give some prominent constraint to this leptoquark interpretation of the muon $g-2$ anomaly, we need to consider other more stringent constraints. We find that the contact coupling $\mu\mu tt$ also depends on $\lambda^{\prime 2}/m_{X_1}^2$, so that the $(g-2)_\mu$ contribution cannot escape from its constraints, which mainly come from the measurement of $Z \to \mu^+ \mu^-$. As a result, the constraints for this contact operator listed in Refs.~\cite{Carpentier:2010ue,Davidson:2011zn} can be translated to the following bound,
\begin{align}\label{}
 \frac { \lambda '^{ 2 } }{ { m }_{ { X }_{ 1 } }^{ 2 } } &\lesssim \frac { 4{ G }_{ F } }{ \sqrt { 2 }  } \times0.07\sim 2.3\times { 10 }^{ -6 }{\rm GeV }^{ -2 } ,
\end{align}
which leads to the loose limit on the $\delta a_\mu$ from the present leptoquark $X$
\begin{align}\label{}
\delta { a }_{ \mu  }&\lesssim2.53\times10^{-6}.
\end{align}

Another relevant constraint to our discussion of $(g-2)_\mu$ is from the muon electric dipole moment (EDM), $d_\mu$. 
Currently,
% the Muon ($g-2$) Collaboration gave the most precise measurement of $d_\mu$ with 
the upper bound of $\left| { d }_{ \mu  } \right|$ is  $1.9 \times { 10 }^{ -19 }~ e\cdot {\rm cm}~(95\% ~{\rm C.L.})$~\cite{Bennett:2008dy}, which will be  improved to be at the level of ${\cal O}(10^{-24}) e\cdot{\rm cm}$ by the J-PARC New $g-2$/EDM experiment Collaboration~\cite{Iinuma:2011zz} in the near future. On the other hand, $d_\mu$ arises dominantly from the top-leptoquark loop in the present leptoquark model, 
given by,
\begin{align}
\left| { d }_{ \mu  } \right| \simeq \frac { e{ m }_{ t } }{ { 16\pi  }^{ 2 }{ m }_{ X_1 }^{ 2 } } f\left(\frac { { m }_{ t }^{ 2 } }{ { m }_{ X_1 }^{ 2 } } \right)\left| {\rm Im}\left[ { \tilde { \lambda  }  }_{ e }^{ 23 }{ \tilde { \lambda  }  }_{ u }^{ 32 } \right]  \right|\,,
\end{align} 
where $f(x)$ is defined in Eq.~(\ref{f-function}). 
Since the leptoquark contribution to $\mu \to e\gamma$ depends on the same function $f(x)$, we can express the coupling bound from 
$d_\mu$ in terms of ${\cal B}(\mu\rightarrow{e}{\gamma})$,
\begin{align}
\frac { \left| {\rm Im}\left[ { \tilde { \lambda  }  }_{ e }^{ 23 }{ \tilde { \lambda  }  }_{ u }^{ 32 } \right]  \right|  }{ \lambda  } \sqrt { {\cal B}(\mu \rightarrow e\gamma ) } \lesssim 36.2\, .
\end{align}
In order to show the constraining power, if the leptoquark gave a branching ratio equal to the current experimental bound of ${\cal B}(\mu \rightarrow e\gamma )<5.7\times 10^{-13}$, the muon EDM would constrain the couplings to be 
\begin{align}
\frac { \left| {\rm Im}\left[ { \tilde { \lambda  }  }_{ e }^{ 23 }{ \tilde { \lambda  }  }_{ u }^{ 32 } \right]  \right|  }{ \lambda  } \lesssim 4.8\times { 10 }^{ 7 }.
\end{align}
Note that $d_\mu$ only relates to the imaginary part of  $\tilde {\lambda^{23}_{e}}\tilde{\lambda^{32}_{u}}$, while $(g-2)_\mu$ its real part. Thus, the muon EDM cannot affect our general conclusion on  $(g-2)_\mu$. 
%The same method can give us the similar approximate leptoquark contribution to the electron EDM
Similarly, one has
\begin{align}
\left| { d }_{e  } \right| \simeq \frac { e{ m }_{ t } }{ { 16\pi  }^{ 2 }{ m }_{ X_1 }^{ 2 } } f\left(\frac { { m }_{ t }^{ 2 } }{ { m }_{ X_1 }^{ 2 } } \right)\left| {\rm Im}\left[ { \tilde { \lambda  }  }_{ e }^{ 13 }{ \tilde { \lambda  }  }_{ u }^{ 31 } \right]  \right|,
\end{align}
which would lead to 
%and the bound to the couplings 
${ \left| {\rm Im}\left[ { \tilde { \lambda  }  }_{ e }^{ 13 }{ \tilde { \lambda  }  }_{ u }^{ 31 } \right]  \right|  }/{ \lambda  } \lesssim 2.2\times { 10 }^{ -2 }$ 
with the most recent upper limit of $|d_e|<8.7\times 10^{-29}~ e\cdot cm$ by 
the ACME Collaboration~\cite{Baron:2013eja}. However, due to the different dependence of the Yukawa couplings, the 
limit from $d_e$  cannot place  a meaningful constraint on the leptoquark solution to the $(g-2)_\mu$ anomaly and $\mu \to e\gamma$.

Therefore, we conclude that this leptoquark model is promising to solve the $(g-2)_\mu$ anomaly with the non-universal Yukawa couplings to suppress $\mu \to e \gamma$, even if we further consider other low-energy experimental constraints.

\section{Conclusions}
Motivated by the long-standing $(g-2)_\mu$ anomaly and the recently updated $\mu\rightarrow{e}{\gamma}$ upper bound, we have examined the correlations between $\mu\rightarrow{e}{\gamma}$  and $(g-2)_\mu$. The general EFT analysis tells us that it is difficult in explaining the muon $g-2$ anomaly while satisfying the $\mu\rightarrow{e}{\gamma}$ bound in a natural theory with ${\cal O}(1)$ Wilson coefficients, since the cutoff scale obtained by fitting the required $(g-2)_\mu$ discrepancy predicts an unbearably large $\mu \to e \gamma$ rate. After compiling all of the one-loop diagram formulae for the $(g-2)_\mu$ corrections, we have proposed two promising methods to eliminate this tension between the new physics contributions to $(g-2)_\mu$ and $\mu \to e \gamma$: the GIM mechanism and the non-universality of couplings. For the latter method, a leptoquark model has been illustrated as a simple example. As expected, with the appropriate choice of leptoquark Yukawa couplings, it is possible to achieve the goal to understand the $(g-2)_\mu$ anomaly while keeping an experimentally allowed $\mu \to e \gamma$ branching ratio.

As discussed in our leptoquark model, even if the new physics accounting for the required $\delta a_\mu$ can escape the $\mu \to e \gamma$ bound with our two methods, we still need to check other constraints, such as the contact interactions, the unitarity of the CKM and/or PMNS matrices, and so on, especially for those with the same coupling dependence as the explanation of the $(g-2)_\mu$ anomaly. Finally, we hope that our investigation on the possible relation between the muon anomalous magnetic moment and the decay $\mu \to e \gamma$ may shed light on the structure of new physics.

\begin{acknowledgments}
We would like to thank Prof. Svjetlana Fajfer for pointing out a mistake in our previous version. The work was supported in part by National Center for Theoretical Sciences, National Science
Council (NSC-101-2112-M-007-006-MY3) and National Tsing Hua
University (103N2724E1).
\end{acknowledgments}

\end{document}